\documentclass[twocolumn,tightenlines,floatfix]{revtex4}

\usepackage{epsfig} 
\usepackage{amssymb} 
\usepackage{amsmath} 
\usepackage{amsfonts}

\newcommand{\be}{\begin{equation}}
\newcommand{\ee}{\end{equation}}

\newcommand{\epem}{$e^+e^-$}

\begin{document}

\title{Cross sections and energy loss for lepton
pair production in muon transport}

\author{Alexander Bulmahn and Mary Hall Reno}
\affiliation{Department of Physics and Astronomy, University of Iowa, Iowa City, IA 52242}

\begin{abstract}
We reevaluate \epem\ pair production from electromagnetic interactions of muons in transit through materials.
Our approach, through the use of structure functions for inelastic
and elastic scattering and including hadronic recoil, make the formalism
useful for tau pair production at high energies. Our results for
\epem\ pairs agree well with prior evaluations. Tau pair production, however, has significant contributions from inelastic scattering in
addition to the usual coherent scattering with the nucleus and scattering
with atomic electrons.
\end{abstract}

\maketitle

\section{Introduction}

Atmospheric muons, the muons produced by cosmic ray interactions in the 
Earth, are detected by many underground detectors. 
Downward-going muons are a large background
to neutrino induced events
\cite{amanda,icecube,macro,antares}. For a range of energies, measurements of muon fluxes test our understanding
of the underlying interactions that produce the muons and offer the opportunity to test models of
the cross section for cosmic ray interactions with air nuclei.

Measurements of the energy dependence of the atmospheric lepton fluxes 
rely on knowledge of the muon energy loss as a function of distance. Underground and
underwater detectors can effectively probe atmospheric muon energies by looking at the 
muon flux as a function of zenith angle $\theta$. Vertical muons travel a depth $D$, while 
muons incident at angle $\theta$ travel a distance $\sim D/\cos\theta$. The electromagnetic
energy loss is essential to the unfolding process between detected and surface
muon fluxes \cite{bugaev,kbs}.

In this paper, we reevaluate the cross section for muon-atom scattering to produce lepton pairs, and
the energy loss of muons from the production of electron-positron pairs.
The largest contribution to the muon energy loss parameter $\beta$ in the energy loss formula
\begin{equation}
\langle \frac{dE}{dX}\rangle = -(\alpha + \beta E) 
\end{equation}
comes from pair production \cite{lkv,ls,lbl,tannenbaum}.
The ``pair-meter'' method of muon energy determination also relies
on the \epem\ pair production cross section \cite{pairmeter,pairmeter1,pairmeter2}.

Evaluations of muon cross sections and energy loss from pair production in collisions
of a muon and a static nucleus have a long history 
\cite{racah,mott,kelner,kk,kp,tsai,abs,Kelner:1998mh,Kelner:2001fg,henry,tannenbaum1,tridents,abgs,Kelner:2000va}.
In this paper, we present the cross section in terms of form factors and structure functions \cite{tsai}
that are not restricted to low lepton masses or low momentum transfers to the target.
We include target recoil in the kinematics \cite{abs}, and
the inelastic contribution \cite{tridents,Kelner:2001fg} to the cross section and energy loss parameter
$\beta_{\rm pair}$ using structure functions parameterized from HERA data \cite{Abramowicz,allm}. 
Our results are compared with the commonly used parameterization of Kokoulin and Petrukhin \cite{kp}.
We do not address the production of muon pair because of the extra contributions from having
identical particles in the final state \cite{henry,tannenbaum1,tridents,abgs,Kelner:2000va}.

We begin by reviewing the calculation in a formalism applicable to elastic and
inelastic scattering including the hadronic recoil. We show the standard
generalization to atoms.
Our results for cross section and energy loss parameter
$\beta$ for protons, hydrogen and higher $Z$ atoms
are shown in Sec. III. In Sec. IV we show an application to high energy \epem\
and $\tau^+\tau^-$ pairs.
Many of the calculational details are relegated to the appendix.

\section{Notation and formalism}

Our organization of the matrix element squared follows the notation of Kel'ner in Ref. \cite{kelner},
however, here we include target recoil.
Akhundov, Bardin and Shumeiko evaluated lepton pair production in
$\mu p$ scattering with proton recoil in Ref. \cite{abs}. We
follow their notation for the kinematics. 

For definiteness, we begin with $e^+e^-$ pair production.
The incident muon with four-momentum $k$ interacts
with a nucleus of four-momentum $p$. The outgoing muon ($k_1$), electron ($k_2$) and positron ($k_3$)
are accompanied by a hadronic final state with $p_H = \sum_{x={\rm hadrons}} p_x$. Appendix A
includes the definition of relevant Lorentz scalars. Note that $p^2=M_t^2$ is the target mass squared, $k^2=k_1^2=m_\mu^2$ and
$k_2^2=k_3^2=m_e^2$. When the target is a proton or a neutron, we set
$M_t=M$.

\begin{figure}%
\includegraphics[width=2.5in]{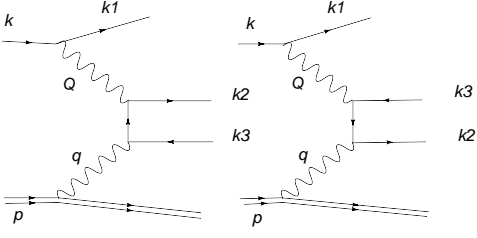}%
\caption{The dominant contribution in $\mu p$ production of $e^+ e^-$
pair production comes from virtual photon graphs shown here. The figure was produced using
Jaxodraw \cite{jaxodraw}.}%
\label{fig:dominant}%
\end{figure}

There are a number of diagrams that contribute to \epem\ pair production.
For $\mu p$ scattering, there are diagrams
where a virtual photon radiated from the muon or proton splits directly into an $e^+e^-$ pair.
The dominant contributions to the pair production cross section come from the two diagrams shown in
Fig. \ref{fig:dominant} \cite{kk,abgs,tridents}. 
As a simplification, we include only the diagrams in Fig. \ref{fig:dominant} 
in our evaluation.

The strategy to evaluate the differential cross section is outlined in Kel'ner \cite{kelner}, and
extended here to include the inelastic case including recoil of the final state hadrons.
The hadronic matrix element squared $H^{\mu\nu}$ is related to a decomposition into structure functions
$F_1$ and $F_2$ which depend on $q^2 = (p-p_x)^2\equiv -t$ and $x_{Bj} = q^2/2p\cdot q$ \cite{devenish}:
\begin{eqnarray*}
e^2H^{\mu\nu} &=& \frac{1}{2}\sum_{{\rm spins},\ X} \langle X| J^\mu | p\rangle \langle X|J^\nu|p\rangle^*\\
W^{\mu\nu} &\equiv & \frac{1}{4\pi}\int H^{\mu\nu} (2\pi)^4\delta ^4(p - q - \sum p_x) \\
&\cdot& \prod_x \frac{{\rm d}^3 p_x}{2E_x (2\pi)^3}\\
&=& -g^{\mu\nu}{W_1}+ p^\mu p^\nu \frac{W_2}{M_t^2}\\
&=& -g^{\mu\nu} F_1 + \frac{p^\mu p^\nu}{M_t^2}\frac{2 M_t^2 x_{ Bj}}{t} F_2\ .
\end{eqnarray*}
Our choice for the sign of the four-momentum $q$ is opposite that of the usual choice for inelastic scattering with 
protons.

The formalism in terms of $W^{\mu\nu}$ is relevant to both inelastic and elastic scattering. For inelastic
scattering, $F_1$ and $F_2$ are dependent on $x_{Bj}$ and $t$ independently. For elastic scattering,
$x_{Bj}=1$ and $F_1$ and $F_2$ are proportional to the delta function $\delta(t+2p\cdot q)=\delta(M_X^2-M_t^2)$, where we have made
the assignment $(\sum p_x)^2 = M_X^2$.

The spin averaged matrix element squared for the muon part of the diagram is
\begin{equation}
A_{\alpha\beta} =\frac{1}{2} {\rm Tr} (/\llap{k}_1+m_\mu)
\gamma_\alpha (/\llap{k}+m_\mu)\gamma_\beta\ .
\label{eq:aalphabeta}
\end{equation}
The change in muon momentum is defined to be $Q\equiv k-k_1$ with $Q^2 = -Y$.

The $\gamma^* (Q)+ \gamma^* (q) \rightarrow e(k_2) + \bar{e}(k_3)$ matrix element comes from the
two diagrams shown in Fig. 1. The result is represented by $e^4 B^{\alpha\beta}_{\mu\nu}$, so that the
differential cross section can be written as
\begin{eqnarray}
\nonumber
{\rm d}\sigma &=& \frac{1}{2\sqrt{\lambda_s}} A_{\alpha \beta}B^{\alpha\beta}_{\mu\nu} 
\frac{e^8}{t^2 Y^2}\delta^4 (k+q - \sum_{i=1}^3 k_i )\\
& \times & 
\prod_i \frac{{\rm d}^3 k_i}{2E_i (2\pi)^3} 4\pi  W^{\mu\nu}{\rm d}^4 q
\label{eq:dsig}
\end{eqnarray}
where $\lambda_s = (2p\cdot k)^2 - 4 m_\mu^2 M_t^2$.

The details of the phase space integrals are in the Appendix,
however, because of its length,
we have not included an explicit expression for
$B^{\alpha\beta}_{\mu\nu}$.
We have evaluated $B^{\alpha\beta}_{\mu\nu}$ using the symbolic manipulation program FORM \cite{form}. We follow
Kel'ner's calculational simplifications \cite{kelner}
by projecting out different terms in the matrix element squared integrated
over most of the phase space. Ultimately, we evaluate
numerically $d\sigma/dy$, $\sigma$ and
\begin{equation}
\label{eq:beta}
\beta _{\rm pair}\equiv \frac{N_A }{A} \int y\ \frac{d\sigma}{dy}\ dy
\end{equation}
where $y\equiv p\cdot Q/p\cdot k$ is the usual inelasticity parameter and $d\sigma/dy$ is the differential cross section for
$e^+e^-$ pair production. In the target rest frame, $p=(M_t,0,0,0)$,
$y=(E-E_1)/E$, the difference in incoming and outgoing muon energies, normalized by the incoming muon energy. In eq. (\ref{eq:beta}), $N_A$
is Avogadro's number and $A$ is the atomic mass.

The structure functions $F_1$ and $F_2$ carry information about the nucleus as well as about the electron cloud which screens the nucleus. 
We discuss the elastic and inelastic cases separately in the next sections. 
It is convenient to further divide the contributions to the cross section according to 
the dependence on the nuclear charge $Z$.
 
\subsection{Coherent scattering with the screened nucleus}

Coherent scattering with the nucleus amounts to elastic scattering
with the charge $Z$ nucleus of mass $M_t=M_A=AM$, leading to a factor
of $Z^2$.
As we show below, for elastic scattering, a delta function enforcing
$M_X^2 = M_A^2$ 
($t=-2p\cdot q$) appears in the structure functions.

For coherent scattering with atom of mass $M_A$, atomic mass $A$ and charge $Z$, the structure functions in the standard approach
are written in terms of 
$\tau\equiv {t}/({4M_A^2})$, 
\begin{eqnarray}
F_1^{\rm coh} &=& \frac{t}{2}G_M^2\ \delta (M_X^2 - M_A^2)\ ,\\
F_2^{\rm coh} &=& \frac{t}{1+\tau}\Bigl( (G_E - F_e)^2
+\tau G_M^2\Bigr)\ \delta (M_X^2-M_A^2) .
\end{eqnarray}
The electric form factor $G_E(t)$ accounts for the electric charge distribution
in the nucleus, with $G_E(0) = Z$, whereas $F_e(t)$ accounts for the $Z$ 
electrons in the atom which screen the nucleus at large distances.

We begin with the nucleon form factors, following with nuclear form factors.
A recent review of the elastic form factors appears in 
Ref. \cite{Perdrisat:2006hj}. Traditionally, the electric and magnetic
form factors ($G_E$ and $G_M$) are written in dipole form. For the
proton
\begin{eqnarray}
\label{eq:dipole}
G(t) &=& \frac{1}{(1+t/0.71\ {\rm GeV}^2)^2}\\
G_{Ep} &=& G(t)\\
G_{Mp} &=& \mu_P\, G(t) = 2.79\, G(t)\ .
\end{eqnarray}

The neutron form factors are
\begin{eqnarray}
G_{En} &=& -\frac{\mu_N\tau }{1+5.6\tau} G(t)
= \frac{1.91\tau }{1+5.6\tau}G(t)\\
G_{Mn} &=& \mu_N\,  G (t) = - 1.91  \, G(t) \ .
\label{eq:neutronmag}
\end{eqnarray}
These results reproduce reasonably well the form factors extracted
by the Rosenbluth separation method. 
Data from 
studying the polarization transfer of polarized electrons to proton targets have yielded a different set of form factors \cite{Perdrisat:2006hj}. For the 
total cross section and energy loss parameters, the new parameterizations
yield essentially the same results as Eqs. (6-10) above.

The electric nuclear form factor \cite{hofstadter}, for large $A$, 
can be represented by \cite{ab,tsai}
\begin{eqnarray}
G_E (t) & = & \frac{Z}{(1+a^2t/12)^2}\\
a &=& (0.58+0.82 A^{1/3})5.07\ {\rm GeV}^{-1}\ .
\label{eq:genuc}
\end{eqnarray}
For large $A$, we set $G_M\simeq 0$. This is a reasonable approximation
because, roughly, the net magnetic moment of a multi-nucleon atom is
small, and the prefactor of $G_M$ in the cross section is also small.

The electric form factor associated with electron screening for hydrogen is obtained by using the electronic wave function to determine the charge density \cite{hofstadter}. This leads to \cite{tsai}
\begin{equation}
F_e(t) = \frac{1}{(1+a_0^2t/4)^2} \quad (Z=1)\ .
\label{eq:fe1}
\end{equation}
Here, $a_0$ is the Bohr radius, $a_0=137/m_e$.
For higher charge and atomic numbers (beyond helium), 
an approximate parameterization of
the electronic electric form factor is \cite{tsai,schiff}
\begin{eqnarray}
F_e (t) & = & \frac{Z}{1+a_e^2t}\ \quad (Z>2)\\
a_e &=& 111.7 Z^{-1/3}/m_e .
\label{eq:fez}
\end{eqnarray}
For large Z, the dominant contribution to the elastic cross section
is proportional to $(G_E-F_e)^2\sim Z^2$ at short distances and large
$t$, however,
at large distances, $(G_E-F_e)^2\sim 0$, where the nucleus is completely
screened.

\subsection{Incoherent scattering with nucleons and electrons}

One component of the cross section for 
incoherent scattering to produce charged lepton pairs comes from
elastic scattering with the 
individual protons and neutrons in
the nucleus. Here, the target mass is $M_t=M$, and one parameterization
of the structure functions gives
\begin{eqnarray}
\nonumber
F_1^{\rm incoh,N} &=& C(t)\frac{t}{2}(ZG_{Mp}^2 +(A-Z) G_{Mn}^2)\\
&& \times \delta (M_X^2-M^2)\\
F_2^{\rm incoh,N} &=& C(t)\frac{t}{1+t/4M^2}\Bigl( ZG_{Ep}^2 +(A-Z)G_{En}^2\\
\nonumber 
&+&\frac{t}{4M^2} (ZG_{Mp}^2+(A-Z)G_{Mn}^2)\Bigr)\ \delta (M_X^2-M^2)\ .
\end{eqnarray}
The prefactor $C(t)$ is the Pauli suppression factor. Following Tsai \cite{tsai}
\begin{equation}
C(t) = \begin{cases}
 \frac{3Q_P}{4P_F}\Bigl( 1-\frac{Q_P^2}{12 P_F^2}\Bigr) & Q_P<2P_F\\
 1 & {\rm otherwise}
\end{cases}
\end{equation}
where $Q_P^2 = t^2/(4M^2) + t$ and $P_F=0.25$ GeV.

There is scattering with the atomic electrons themselves \cite{ab}.
For scattering with electrons, the target mass goes from $M_A$ or
$M$ to $m_e$, and
\begin{eqnarray}
F_1^{\rm incoh,e} &=& \frac{t}{2}Z\ \delta (M_X^2-m_e^2)\\
F_2^{\rm incoh,e} &=& \frac{t}{1+t/4m_e^2}Z\Bigl( F_e^{\rm incoh}(t)\\
\nonumber
&+&\frac{t}{4m_e^2}\Bigr) \delta (M_X^2-m_e^2)\ .
\end{eqnarray}
For the hydrogen atom,
$F_e^{\rm incoh} = 1-F_e(t)^2$ with $F_e(t)$ given by eq. (\ref{eq:fe1}).  
We use the parameterization of Ref. \cite{ab} for higher $Z$ atoms, where 
\begin{eqnarray}
F_e^{\rm incoh} &=& \frac{ c^4 t^2}{(1+c^2t)^2}\\
c &=& 724 \ Z^{-2/3}/m_e
\label{eq:elinel}
\end{eqnarray}
For the high energies of interest here, neglecting addition diagrams coming from identical particle exchange in $\mu e^-\rightarrow \mu e- e^+ e^-$ is an acceptable approximation \cite{Kelner:1998mh}.

\subsection{Inelastic scattering with nucleons}

For $\mu A$ scattering where the momentum transfer is large enough that we are above threshold for pion production,  inelastic scattering accounting for the substructure of the nucleons is required. For the proton structure function
$F_2^p(x_{Bj},t)$,
we use the Abramowicz, Levin, Levy and Maor\cite{Abramowicz} (ALLM) parameterization, updated in
Ref. \cite{allm}. We do not include the parameterization here, but refer the reader to Ref. \cite{allm}. The parameterization also appears in
an appendix of Ref. \cite{Dutta}. This parameterization of the structure function
is valid over the important range of small $t$, the dominant
contribution to the inelastic cross section. It agrees well with data over a wide range of $x_{Bj}$ and $t$
including the perturbative regime.

For inelastic scattering with a nuclear target rather than a 
proton target, the process
is still probing the structure of individual nucleons, so $M_t=M$.
There is a nuclear effect that modifies the proton structure functions, called nuclear shadowing, which we incorporate with \cite{shadow}
$a(A,x_{Bj},t)\simeq a(A,x_{Bj})$ and
\begin{equation}
a(A,x) = 
\begin{cases}
A^{-0.1} & x_{Bj}<0.0014\\
A^{0.069\log_{10}x_{Bj}+0.097} & 0.0014<x_{Bj}<0.04\\
1 & 0.04<x_{Bj}\ .
\end{cases}
\end{equation}
The nuclear structure functions are taken to be
\begin{eqnarray}
F_2^A &=& a(A,x_{Bj}) (Z + (A-Z)P(x_{Bj}))F_2^p\ \\
F_1^A &=& F_2^A/(2x_{Bj})
\end{eqnarray}
where $P(x_{Bj}) = 1 - 1.85x_{Bj}+2.45x_{Bj}^2-2.35x_{Bj}^3+x_{Bj}^4$ accounts for the difference between proton and
neutron structure functions \cite{neutrons}.

\section{Results}

\subsection{Cross sections}
\begin{figure}
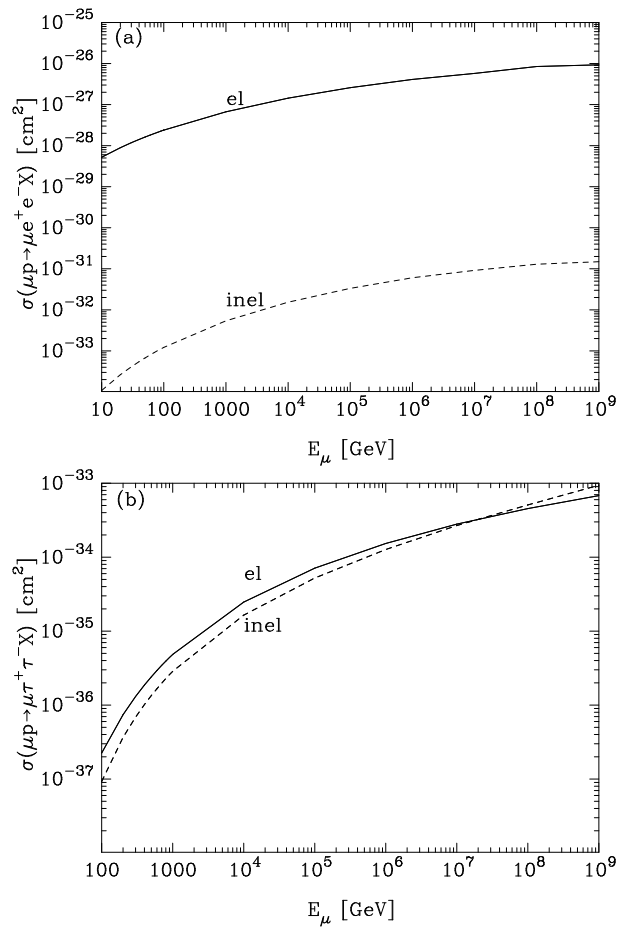
%
\includegraphics[width=2.4in,angle=270]{sig-elp-fig2.ps}
\includegraphics[width=2.4in,angle=270]{sig-taup-fig2.ps}
\caption{Cross section for $\mu p\rightarrow \mu e^+ e^- X$ (a)
and $\mu p\rightarrow \mu \tau^+ \tau^- X$ (b), 
showing the elastic and
inelastic contributions separately.}%
\label{fig:sigp}%
\end{figure}

We begin with muon scattering from protons to produce a pair of charged
leptons. The cross section is the sum of the cross section for $\mu p$ elastic scattering using the proton form
factors in eqs. (\ref{eq:dipole}-\ref{eq:neutronmag})
and the inelastic scattering term:
\begin{equation}
\sigma _{\mu p} = \sigma_{p}^{\rm coh}+\sigma_{p}^{\rm inel}\ .
\end{equation} 
For $e^+ e^-$ pair production shown in Fig. \ref{fig:sigp}a, 
the elastic scattering is $\sim 5-6$ orders of magnitude larger than
the inelastic contribution. This can be understood by noting that the phase
space integral with the photon propagator involves $dt/t^2$, making $t$ near
the minimum $t$ the dominant part of the integral. The minimum $t$
for elastic scattering is  
$$t_{\rm min}\simeq \frac{4 m_e^2 m_\mu^2}{E^2}\ ,$$
as discussed in detail in, for example, Appendix A of Ref. \cite{tsai}.
The structure function
$F_{2,p}^{\rm inel}$ is small for $t\ll M$.

The relevant scale for $t$ is much larger for $m_e\rightarrow m_\tau$ in
$\tau^+\tau^-$ pair production. Fig. 
\ref{fig:sigp}b shows that the inelastic contribution is comparable to 
the elastic cross section for a range of energies. The inelastic cross section contributes between 30\% and 60\% of the total cross section for
muon energies between $10^2-10^9$ GeV. Overall, however, the tau pair
cross section is significantly smaller than the electron-positron pair production cross section. Tau pair production is not an important component of muon energy loss. Muon pair production is intermediate
between the two sets of cross sections in Fig. \ref{fig:sigp}
\cite{tridents}.
As noted above, we do not include muon pair production here due to the additional exchange diagram required by Fermi statistics for the identical fermions. 

\begin{figure}
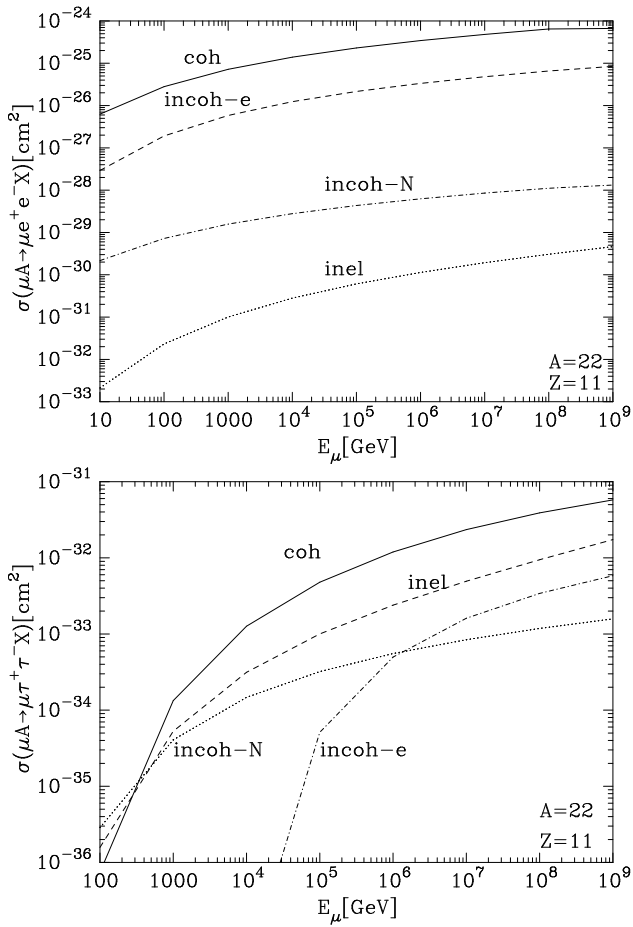
%
\includegraphics[width=2.4in,angle=270]{sigma-rock.ps}
\includegraphics[width=2.4in,angle=270]{sigma-rock-taup.ps}
\caption{Cross section for (a)$\mu A\rightarrow \mu e^+ e^- X$ and (b)$\mu A\rightarrow \mu \tau^+ \tau^- X$
for standard rock ($Z=11$ and $A=22$), for coherent, incoherent and inelastic contributions.}
\label{fig:sigrock}
\end{figure}

The targets of interest for atmospheric muons or muons produced by neutrinos are atoms. Fig. \ref{fig:sigrock}a shows the contributions to the cross section from coherent scattering, incoherent scattering with electrons and nucleons and inelastic scattering, for
$e^+ e^-$ pair production by muons on a rock target, using the standard
rock values of $A=22$ and $Z=11$. Fig. \ref{fig:sigrock}b shows the same quantities for tau pair
production.

For $e^+e^-$ production, the coherent contribution dominates,
followed by incoherent scattering on protons and electrons. Incoherent scattering with neutrons is at the level
of $\sigma\sim 10^{-32}-10^{-30}$ over the range of incident muon energies from
$10-10^9$ GeV. We note that 
our result for the incoherent scattering with atomic electrons
agrees with the approximate analytic formula of Kelner in eq. (46) of
Ref. \cite{Kelner:1998mh} to within 2\% at $E_\mu=100$ GeV, and is about
18\% larger at $E_\mu=10^9$ GeV for $Z=11$. Our result for incoherent scattering with nucleons is
a factor of $\sim 10$ larger than in Ref. \cite{ab}, in which the form factor is different. However, the consequences for \epem\ pair production does not depend significantly on the incoherent nucleon contribution.

As in the case with proton targets, the inelastic contribution
to \epem\ pair production
is orders of magnitude smaller than the elastic contribution.
Tau pair production has a different balance of contributions. The coherent cross section
dominates the total cross section to a lesser degree. As the muon energy increases well beyond the threshold for production of tau pairs in scattering with electrons, incoherent scattering with electrons becomes increasingly important. Inelastic scattering contributions are important,
especially at the lower energies.
Fig. \ref{fig:sigrock}b shows the threshold energy dependence of the cross section for incoherent scattering
with electrons.
The Pauli suppression factor is particularly relevant in the evaluation of incoherent scattering with nucleons at high energies. Our result for incoherent scattering with nucleons is a factor of 
$\sim 2-4$ lower for tau pair production than the cross section coming from  
using the form factor in Ref. \cite{ab}.

\begin{figure}
\includegraphics[width=2.4in,angle=270]{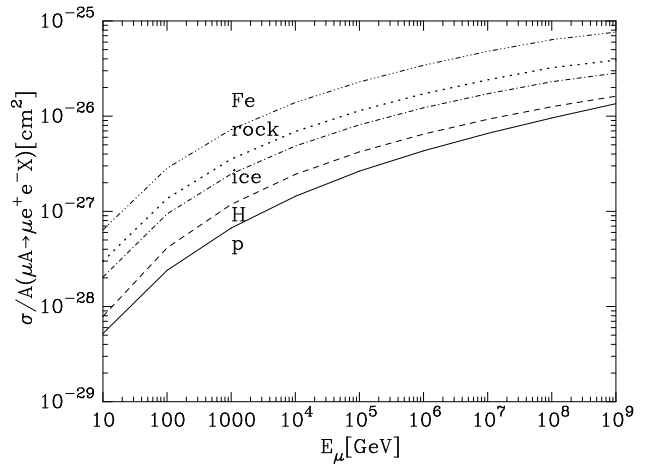}%
\caption{Cross section  divided by atomic number $A$ for $\mu A\rightarrow \mu e^+ e^- X$, for $A=1$ (proton and hydrogen), $A=14.3$ (ice), $A=22$ (standard rock), $A=55.9$ (iron).}
\label{fig:sigoA}
\end{figure}

The results for other targets are shown in
Fig. \ref{fig:sigoA}, where the total $e^+e^-$ pair production
cross section is divided by $A$.

\subsection{Energy loss parameter $\beta_{\rm pair}$}

\begin{figure}%
\includegraphics[width=2.4in,angle=270]{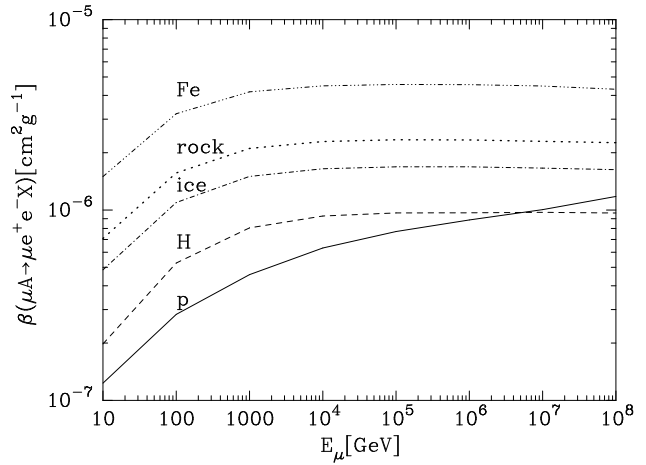}%
\caption{Energy loss parameter $\beta_{\rm pair}$ via $e^+ e^- $ pair production
for proton, hydrogen, ice, rock and iron.}%
\label{fig:beta}%
\end{figure}

The energy loss parameter $\beta_{\rm pair}$ is defined in eq. (\ref{eq:beta}). This parameter is shown in Fig.
\ref{fig:beta} for a variety of materials. For protons, $\beta_{\rm pair}$ rises with energy, however, for
atomic targets with high energy muons, the atomic screening of the nucleus at large distances cuts off the growth of the parameter at high energies. The contribution to $\beta_{\rm pair}$ from tau pair production is suppressed by at least four orders of magnitude, depending on the muon energy.

Our evaluation of $\beta_{\rm pair}$ compares well with the analytical form of Kokoulin and Petrukin (KP) \cite{kp,lkv}. At $E_\mu = 10 $ GeV for
rock, our result is 2.5\% lower for $\beta_{\rm pair}$ than the KP evaluation. At 100 GeV, the two results agree to within $<1\%$. 
At $E_\mu = 10^8$ GeV, our evaluation gives $\beta_{\rm pair}$ lower
by $\sim 4\%$. In the energy range of $10-100$ GeV, our energy loss parameter is $2-3\%$ larger than that of Groom, Mokhov and Striganov
\cite{lbl}. At $E=10^5$ GeV, our results agree to within the numerical
accuracy of our evaluation ($<0.1\%$).

\section{Discussion} 

One advantage of our approach is that we are not confined to the low $t$ or small lepton mass limits. 
Although this is the dominant limit for the total cross section and the energy loss parameter for
\epem pair production, 
one can make interesting use of the formalism to consider high energy charged lepton pairs. 
The lepton pair energy (in the target rest frame) is
\begin{equation}
E_{\rm pair} = \frac{T+S_x}{2M_t}
\end{equation}
in terms of invariants defined in the Appendix.
Figs. \ref{fig:sigice}a and \ref{fig:sigice}b 
show the total cross sections for
lepton pair production by muons in ice,
when the total pair energy is larger than 50 GeV. 

As in the case of the total cross section, for $e^+e^-$ pair production,
the dominant contribution to the high energy pair cross section is coherent
scattering with the nucleus, with a $\sim 10\%$ correction from incoherent
scattering with the atomic electrons.
The cross section for producing \epem with $E_{\rm pair}$=50 GeV
for $E_\mu\sim 10^3$ GeV is $\sigma \sim 10^{-27}$ cm$^2$, equivalent
to an interaction length of $\sim 20$ m. With the potential to measure
electrons at this energy in IceCube, this may be an interesting reducible
background to $\nu_e\rightarrow e$ conversions to electrons.

Tau pair production by 100 GeV muons
has nearly equal contributions from 
coherent scattering, inelastic scattering and incoherent scattering with nucleons. At high energies $E\sim 10^9$ GeV, the inelastic cross section is approximately half of the coherent cross section for 50 GeV tau pair 
production. Again, incoherent scattering with atomic electrons
is about 10\% of the cross section for coherent scattering when the muon energy is well above the threshold for tau pair production.

Tau neutrinos will come
from the decays of tau pairs produced by muon transit through rock or
ice.
Tau neutrino production in the atmosphere is quite low, especially
at low energies, because it requires heavy quark (charm or b quark)
production \cite{pr,stasto}. Neutrino oscillations over the diameter of
the Earth convert muon neutrinos to tau neutrinos. Tau decays from
$\mu A\rightarrow \mu \tau^+\tau^- X$ will contribute to the overall
downward flux of tau neutrinos. Quantitative evaluations of this source
of downward-going tau neutrinos is under investigation.

In summary, we have provided a review of the contributions to
\epem and $\tau^+\tau^-$ production by muons as they interact electromagnetically as they pass through materials. Our approach is flexible, in that it can be applied to high momentum transfers, and to
high mass leptons. Measurements of high energy \epem\ pair production and tau pair (and associated tau neutrino pair) production will test the theoretical underpinnings of this evaluation.

\begin{figure}
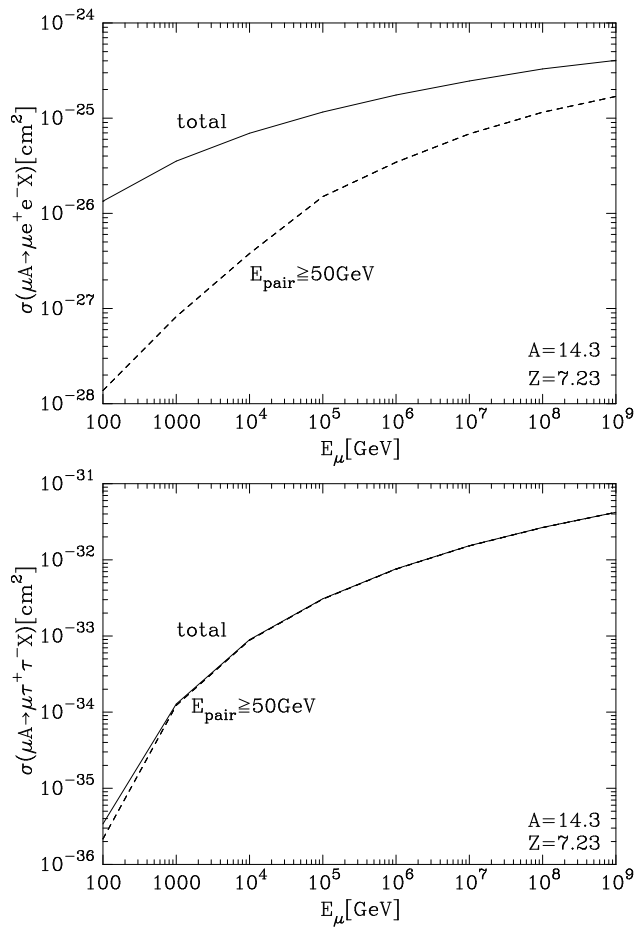

\includegraphics[width=2.4in,angle=270]{sigma-ice-50vstot.ps}
\includegraphics[width=2.4in,angle=270]{sigma-ice-taup-50vstot.ps}%
\caption{Cross section for $\tau^+\tau^-$ pair production and
for $e^+ e^-$ pair production for $E_{\rm pair}>50$ GeV by muon interactions in ice.}
\label{fig:sigice}
\end{figure}

\appendix

\section{Kinematics and cross section}
We present here a compilation of useful kinematic relations and phase
space limits for $e^+e^-$ pair production on a proton target of mass
$M$.

Fig. 1 shows the dominant diagrams for $\mu(k) +P(p)\rightarrow \mu (k_1)
+e(k_2)+\bar{e}(k_3) + X$ where
$$k+p = k_1+k_2+k_3+\sum_x p_x\ .$$
Eq. (\ref{eq:dsig}) shows the differential cross section in terms
of phase space integrals and the quantities
\begin{eqnarray*}
Y& =& -Q^2=-(k-k_1)^2\\
t&=&-q^2= -(p-p_x)^2\\
\lambda_s &=& S^2-4 m_\mu^2 M^2\ .
\end{eqnarray*}

For the evaluation of the phase space, 
it is useful notationally to make the
further definitions of invariants following Ref. \cite{abs}:
\begin{eqnarray*}
V^2&=&\kappa^2= (k_2+k_3)^2 = (q+Q)^2\\
S &=& 2p\cdot k\\
X &=& 2p\cdot k_1\\
S_x &=& 2 p\cdot Q = S-X \\
M_x^2 &=& p_x^2\\
W^2 &=& (p+Q)^2 = M^2+S-X-Y\\
T &=& 2 p\cdot q =M^2-t-M_x^2  \\
\lambda_t &=& T^2+4M^2t\\
\lambda_Y &=& S_x^2 +4M^2 Y\\
\Delta M^2 &=& (M_x^{\rm min}+ 2m_e)^2-M^2
\end{eqnarray*}
Rewriting the differential cross section (eq. (\ref{eq:dsig}))
\begin{equation}
d\sigma
= \frac{1}{2\sqrt{\lambda_s}} A_{\alpha \beta}B^{\alpha\beta}_{\mu\nu} 
\frac{\alpha^4  W^{\mu\nu}}{2\pi^4 t^2 Y^2}\, {\rm d}(PS)\ ,
\end{equation}
the phase space integrals reduce to
\begin{eqnarray*}
{\rm d}(PS) &=&\frac{{\rm d}\phi_1 {\rm d}S_x\, {\rm d}Y {\rm d}V^2\, {\rm d}t\, {\rm d}M_x^2\, {\rm d}\phi_q}
{16\sqrt{\lambda_Y \lambda_s}}{\rm d}\Gamma_{\rm pair}\\
{\rm d}\Gamma_{\rm pair}&=& \delta^4(\kappa - k_2-k_3)\frac{{\rm d}^3 k_2}{2E_2 }
\frac{{\rm d}^3 k_3}{2E_3 }\\
&=& \frac{1}{8}\sqrt{1-4m_e^2/V^2}\ {\rm d}\cos\theta_e\, {\rm d}\phi_e
\end{eqnarray*}

The limits of integration for these integrals are \cite{abs}
\begin{eqnarray*}
(M+m_\pi)^2&\leq & M_x^2\leq (\sqrt{W^2}-2\, m_e)^2\\
4\, m_e^2 &\leq & V^2\leq \frac{1}{2M^2}\Bigl( S_x T+
\sqrt{\lambda_Y}\sqrt{\lambda_t}\Bigr) - t - Y\\
t_{\rm min} &\leq&  t\leq t_{\rm max}\\
t_{\rm min} &=&
(S_x(W^2-M_x^2)+2YM_x^2 - 4 m_e^2(S_x+2M^2)\\
&-&\sqrt{U})/(2W^2)\\
t_{\rm max} &=&
(S_x(W^2-M_x^2)+2YM_x^2 - 4 m_e^2(S_x+2M^2)\\
&+& \sqrt{U})/(2W^2)\\
U&=& \lambda_Y \bigl(W^4+M_x^4+16m_e^4\\
&-& 2(W^2(M_x^2+4m_e^2)+4m_e^2 M_x^2)\bigr)\\
Y^{\rm min} &\leq& Y\leq Y^{\rm max}\\
Y^{\rm min} &=& \frac{\lambda_s - S S_x}{2M^2}\\
&-&\frac{1}{2M^2}\sqrt{(\lambda_s-SS_x)^2 - 4 m_\mu^2 M^2 S_x^2}\\
&=& \frac{\lambda_s - S S_x}{2M^2}
-\frac{1}{2M^2}\sqrt{\lambda_s(\lambda_s - 2SS_x +S_x^2)}\\
Y^{\rm max} &=& S_x - \Delta M^2\\
S_x^{\rm min} &=& \Bigl[\lambda_s + \Delta M^2(S+2M^2) \\&-&
\sqrt{\lambda_s(\lambda_s-2S\Delta M^2+(\Delta M^2)^2 - 4 m_\mu^2\Delta M^2)}\Bigr]\\
&\times & (2(S+m_\mu^2 + M^2))^{-1}\\
S_x^{\rm max} &=& S - 2 M m_\mu\ 
\end{eqnarray*}
Our numerical evaluation of these integrals was performed using the Fortran program VEGAS \cite{lepage}.

The integrals above are for inelastic scattering with a proton,
hence the minimum $M_x^{2,{\rm min}}=(M+m_\pi)^2$. 

Elastic scattering
is enforced by a delta function of the form
$\delta (M_x^2-M_t^2)$ where $M_t$ is the target mass. 
In this case,
it is useful to rewrite the phase space integral involving $q=p-p'$
(for outgoing target momentum $p'$)
and $\kappa = k_2+k_3$ as
$$d^4 \kappa \delta^4(Q+q-\kappa ) d^4 q =
\frac{1}{4\lambda_Y^{1/2}} \, dV^2\, dt\, d(2p\cdot q)\ d\phi\ .$$
In the remaining
integrals, for elastic scattering of a target of mass $M_t$, one makes
the replacement $M\rightarrow M_t$. The target mass also appears in
$\lambda_s$ in eq. (\ref{eq:dsig}).

We note that $q=p-p_x$ is the opposite sign to the usual convention. 
The Bjorken $x$ value, $x_{ Bj}=q^2/(2p\cdot q)$ in these variables is
$$x_{Bj} = \frac{t}{t+M_x^2-M^2}= -\frac{t}{T}\ .$$
 
\begin{acknowledgments}
This research was supported in part by 
US Department of Energy 
contracts DE-FG02-91ER40664. We thank F. Coester for helpful conversations.

\end{acknowledgments}



\begin{thebibliography}{99}
\bibitem{amanda}
T.~DeYoung for~the~IceCube~Collaboration,
  J.\ Phys.\ Conf.\ Ser.\  {\bf 136}, 022046 (2008)
  [arXiv:0810.4513 [astro-ph]].
\bibitem{icecube}
A. Achterberg {\it et al.} [IceCube Collaboration],
Phys. Rev. {\bf D 76}, 027101 (2007).
\bibitem{macro}
  M.~Ambrosio {\it et al.}  [MACRO Collaboration.],
  Phys.\ Rev.\  D {\bf 52}, 3793 (1995).
\bibitem{antares}
M.~Ageron  [ANTARES collaboration],
  arXiv:0812.2095 [astro-ph].
\bibitem{bugaev}
 See, for example,
 E.~V.~Bugaev, A.~Misaki, V.~A.~Naumov, T.~S.~Sinegovskaya, S.~I.~Sinegovsky and N.~Takahashi,
  Phys.\ Rev.\  D {\bf 58}, 054001 (1998)
  [arXiv:hep-ph/9803488].
\bibitem{kbs}
  S.~I.~Klimushin, E.~V.~Bugaev and I.~A.~Sokalski,
  Phys.\ Rev.\  D {\bf 64}, 014016 (2001)
  [arXiv:hep-ph/0012032].
\bibitem{lkv}
W. Lohmann, R. Kopp and R. Voss, CERN Yellow Report No. EP/85-03.
\bibitem{ls}
P.~Lipari and T.~Stanev,
  Phys.\ Rev.\  D {\bf 44}, 3543 (1991).
\bibitem{lbl}
D. E. Groom, N. V. Mokhov and S. Striganov, Atom. Data Nucl.
Data Tabl. {\bf 78}, 183 (2001).
\bibitem{tannenbaum}
M. J. Tannenbaum, Nucl. Inst. Meth. A300, 595 (1991).
\bibitem{pairmeter}
R. P. Kokoulin and A. A. Petrukhin, Nucl. Inst. Meth. A263, 468 (1988);
Sov. J. Part. Nucl. 21, 332 (1990). 
\bibitem{pairmeter1}
 V.~B.~Anikeev {\it et al.},
{\it Proceedings of the 27th International Cosmic Ray Conference (ICRC 2001)}, Hamburg, Germany, 7-15 Aug 2001, 958 (2001).
\bibitem{pairmeter2}
 R.~Gandhi and S.~Panda,
  JCAP {\bf 0607}, 011 (2006)
  [arXiv:hep-ph/0512179].
\bibitem{racah}
G. Racah, Nuovo Cim. {\bf 16}, 93 (1937).
\bibitem{mott}
N. F. Mott and H. S. W. Massey, {\it The theory of atomic collisions}, Clarendon Press (Oxford) 1965.
\bibitem{kelner}
S. R. Kel'ner, Sov. J. Nucl. Phys. {\bf 5}, 778 (1967).
\bibitem{kk}
S. R. Kel'ner and Yu. D. Kotov, Sov. J. Nucl. Phys. {\bf 7}, 237 (1968).
\bibitem{kp}
R. P. Kokoulin and A. A. Petrukhin, in Proceedings of the XII International Conference on Cosmic Rays (Hobart, Tasmania, Australia,
1971) Vol 6.
\bibitem{tsai}
Y.-S. Tsai, Rev. Mod. Phys. 46, 815 (1974).
\bibitem{abs}
A. A. Akhundov, D. Yu. Bardin and N. M. Shumeiko,
Sov. J. Nucl. Phys. {\bf 32}, 234 (1980). 
\bibitem{Kelner:1998mh}
  S.~R.~Kel'ner,
  Phys.\ Atom.\ Nucl.\  {\bf 61}, 448 (1998)
  [Yad.\ Fiz.\  {\bf 61}, 511 (1998)].
\bibitem{Kelner:2001fg}
  S.~R.~Kel'ner and D.~A.~Timashkov,
  Phys.\ Atom.\ Nucl.\  {\bf 64}, 1722 (2001)
  [Yad.\ Fiz.\  {\bf 64}, 1802 (2001)].
\bibitem{henry}
G. R. Henry, Phys. Rev. {\bf 154}, 1534 (1967).
\bibitem{tannenbaum1}
M. J. Tannenbaum, Phys. Rev. {\bf 167}, 1308 (1968).
\bibitem{tridents}
V.~Ganapathi and J.~Smith,
  Phys.\ Rev.\  D {\bf 19}, 801 (1979).
\bibitem{abgs}
A. A. Akhundov, D. Yu. Bardin, N. D. Gagunashvili and N. M. Shumeiko,
Sov. J. Nucl. Phys. {\bf 31}, 127 (1980).
\bibitem{Kelner:2000va}
  S.~R.~Kel'ner, R.~P.~Kokoulin and A.~A.~Petrukhin,
  Phys.\ Atom.\ Nucl.\  {\bf 63}, 1603 (2000)
  [Yad.\ Fiz.\  {\bf 63}, 1690 (2000)].
\bibitem{Abramowicz}
  H.~Abramowicz, E.~M.~Levin, A.~Levy and U.~Maor,
  Phys.\ Lett.\  B {\bf 269}, 465 (1991).
\bibitem{allm}
H. Abramowicz and A. Levy, hep-ph/9712415.
\bibitem{jaxodraw}
D.~Binosi and L.~Theussl,
  Comput.\ Phys.\ Commun.\  {\bf 161}, 76 (2004)
  [arXiv:hep-ph/0309015].
\bibitem{devenish}
  R.~Devenish and A.~Cooper-Sarkar,
{\it  Oxford, UK: Univ. Pr. (2004) 403 p}
\bibitem{form}
J.~A.~M.~Vermaseren,
  arXiv:math-ph/0010025.
\bibitem{Perdrisat:2006hj}
  For a review of recent results, see, 
  e.g., C.~F.~Perdrisat, V.~Punjabi and M.~Vanderhaeghen,
  Prog.\ Part.\ Nucl.\ Phys.\  {\bf 59}, 694 (2007)
  [arXiv:hep-ph/0612014].
\bibitem{hofstadter}
R. Hofstadter, Ann. Rev. Nucl. Sci. {\bf 7}, 231 (1957).
\bibitem{ab}
Yu. M. Andreev and E. V. Bugaev, Phys. Rev. {\bf D 55}, 1233 (1997).
\bibitem{schiff}
L. Schiff, Phys. Rev. {\bf 83}, 252 (1951).
\bibitem{Dutta}
  S.~I.~Dutta, M.~H.~Reno, I.~Sarcevic and D.~Seckel,
  Phys.\ Rev.\  D {\bf 63}, 094020 (2001)
  [arXiv:hep-ph/0012350].
\bibitem{shadow}
E665 Collaboration, M.R. Adams et al., Phys. Rev. Lett. 68, 3266 (1992), Phys. Lett. B
287, 375 (1992), Z. Phys. C 67, 403 (1995).
\bibitem{neutrons}  
A.C. Benvenuti et al. (BCDMS Collaboration), Phys. Lett.
B 237, 599 (1990).
\bibitem{pr}
 L.~Pasquali and M.~H.~Reno,
  Phys.\ Rev.\  D {\bf 59}, 093003 (1999)
  [arXiv:hep-ph/9811268].

\bibitem{stasto}
  A.~D.~Martin, M.~G.~Ryskin and A.~M.~Stasto,
  Acta Phys.\ Polon.\  B {\bf 34}, 3273 (2003)
  [arXiv:hep-ph/0302140].
\bibitem{lepage}
  G.~P.~Lepage,
  J.\ Comput.\ Phys.\  {\bf 27}, 192 (1978).
  
\end{thebibliography}
\end{document}